\documentclass[aip,pof,preprint]{revtex4-1}
\usepackage{amsmath}					
\usepackage{amssymb}					
\usepackage{amsthm}						
\usepackage{cleveref}
\usepackage{isomath}
\usepackage[export]{adjustbox}
\usepackage[english]{babel}
\usepackage[UKenglish]{isodate}
\cleanlookdateon

\newcommand{\Minfty}{\mathcal{M}^\infty}
\newcommand{\R}{\mathcal{R}}

\newcommand{\Rtbexact}{\mathcal{R}^{\text{2B,exact}}}
\newcommand{\Rtbinfty}{\mathcal{R}^{\text{2B,}\infty}}
\renewcommand{\v}[1]{\mathbfit{#1}}
\renewcommand{\t}[1]{{\mathsfbfit{#1}}}
\newcommand{\tu}[1]{\boldsymbol{\mathsf{#1}}}

\begin{document}

\begin{abstract}
In Stokesian Dynamics, particles are assumed to interact in two ways: through long-range mobility interactions and through short-range lubrication interactions. To speed up computations, in shear-driven concentrated suspensions, often found in rheometric contexts, it is common to consider only lubrication. We show that, although this approximation may provide acceptable results in shear-driven, periodic suspensions, for bidisperse suspensions where the particles are exposed to an external force, it can produce physically unreasonable results. We suggest that this problem could be mitigated by a careful choice of particle pairs on which lubrication interactions should be included.
\end{abstract}

\title{Anomalous effect of turning off long-range mobility interactions in Stokesian Dynamics}
\date{\today}
\author{Adam K.\ \surname{Townsend}}
\email{a.townsend@ucl.ac.uk}
\author{Helen J.\ \surname{Wilson}} 
\email{helen.wilson@ucl.ac.uk}
\affiliation{Department of Mathematics, University College London, Gower Street, London WC1E 6BT, UK}
\maketitle

\section{Introduction}\label{sec:introduction}
The Stokesian Dynamics (SD) method \citep{brady_stokesian_1988} of simulating the motion of rigid spherical particles in a Newtonian background fluid is a reliable and flexible approach to modelling low-Reynolds number suspensions. In this regime, the fluid motion is governed by the Stokes equations, and the particle motion is governed by the force and torque balance equations,%
\begin{subequations}%
\begin{align}%
\mathbf{0} &= \v{F}^H + \v{F},\\
\mathbf{0} &= \v{T}^H + \v{T},
\end{align}%
\end{subequations}%
where $\v{F}^H$, $\v{T}^H$ are hydrodynamic forces and torques acting on the particles and $\v{F}$, $\v{T}$ are external forces and torques. If desired, we can also include Brownian forces in this formulation.

The linearity of Stokes flow allows us to relates the suspended particles' first force moments (force $\v{F}$, torque $\v{T}$, stresslet $\t{S}$) to their velocity moments (velocity $\v{U}$, angular velocity $\v{\Omega}$, rate of strain $\t{E}$),
\begin{equation}
	\begin{pmatrix}\v{F}\\\v{T}\\\t{S}\end{pmatrix} = \R \begin{pmatrix}\v{U}\\\v{\Omega}\\\t{E}\end{pmatrix},
	\label{resistance-form}
\end{equation} 
through a `grand resistance matrix'\citep{brenner_stokes_1972}, $\R$.

Although this grand resistance matrix can be generated exactly (for example, with the boundary element method), this is computationally expensive. SD provides a method to generate a good approximation to this grand resistance matrix at much less expense. For particles at large separation distances, Fax\'{e}n's laws \citep{kim_microhydrodynamics_2005} provide asymptotic expressions for the velocity of the surrounding fluid. These come in the inverse (`mobility') form to \cref{resistance-form}, and fill a mobility matrix, $\Minfty$. At short separation distances, the majority of the hydrodynamic force on a particle comes from the strong pressure gradients required to squeeze fluid out from between it and its neighbour. For interacting pairs of spheres we have full expressions\citep{kim_microhydrodynamics_2005} for this lubrication-dominated fluid motion, and by treating all near-field interactions as pairwise, the two-body resistance matrix $\Rtbexact$ is constructed. Since the lubrication expressions already include mobility interactions, to prevent double-counting of each interacting pair, we need to remove their associated mobility interactions. The mobility interaction for each pair is computed as the mobility matrix, $\mathcal{M}^{\text{2B,}\infty}$, for the two-particle system, which is then inverted and placed, for each pair, appropriately in the full system to form another two-body resistance matrix, $\Rtbinfty$. SD combines these far and near regimes to form an approximation to the grand resistance matrix which works well at all separation distances, 
\begin{equation}
	\R_\text{SD} = (\Minfty)^{-1} + \Rtbexact - \Rtbinfty.
	\label{grand-resistance-matrix}
\end{equation}

The lubrication resistance matrices, $\Rtbexact$ and $\Rtbinfty$, are typically sparse, as they are calculated only for pairs of particles which are sufficiently close together, normally with a scaled separation distance less than a critical value, $r^*$ (also $r_c$ in the literature),
\begin{equation}%
	s < \frac{r^*}{2}(a_1 + a_2),%
	\label{r-star}%
\end{equation}%
where the centres of two particles of radius $a_1$ and $a_2$ are a distance $s$ apart. A typical value\citep{banchio_accelerated_2003} for $r^*$ is 4. Meanwhile, the long-range mobility matrix, $\Minfty$, considers the motion of each particle as a result of all other particles, so is always dense.

An technique to avoid computing and inverting $\Minfty$, common in concentrated suspensions of Brownian particles \citep{ball_simulation_1997,bybee_hydrodynamic_2009,kumar_microscale_2010,banchio_accelerated_2003,ando_dynamic_2013} but also seen with non-Brownian particles \citep{torres_large-scale_1996}, is for simulators to assert that although the long-range hydrodynamic interactions decay slowly (like $1/r$), they are screened by the many-body effects in the dense suspension. The effective motion of the particles is governed predominantly by their neighbours rather than the hydrodynamics of the system as a whole. In other words, the large number and strength of near-field lubrication forces exceeds the effect of the far-field hydrodynamic forces. This theory was first established---for entangled polymer solutions---by \citet{de_gennes_dynamics_1976}, who gave a cutoff distance for a given polymer concentration, after which long-range hydrodynamic interactions can be ignored.
  
In this case, researchers replace the dense $\Minfty$ (in both the first and third terms in the right-hand side of \cref{grand-resistance-matrix}) with its far-field limit: a drag-only or `lubrication hydrodynamics' (LH) approximation. This limit, which can be seen from Fax\'{e}n's laws as $r \to \infty$, consists solely of self-terms on the leading diagonal of the matrix. In particular, for identical particles of radius $a$, it is given by
\begin{equation}
	\Minfty_{\text{LH}} =
	\begin{pmatrix}
	\displaystyle\frac{\t{I}}{6\pi\mu a}   & \tu{0}    & \tu{0}        \\
	\tu{0}   & \displaystyle\frac{\t{I}}{8\pi\mu a^3} & \tu{0}        \\
	\tu{0}   & \tu{0}    & \displaystyle\frac{\t{I}}{\frac{20}{3}\pi\mu a^3} 
	\end{pmatrix},
	\label{minfty-far-field-limit}
\end{equation}
where $\t{I}$ is the appropriately-sized identity matrix. The viscosity term, $\mu$, is often replaced with an effective viscosity, $\eta(\phi)$, dependent on suspension concentration, $\phi$. For monodisperse suspensions, this may be chosen to be the dilute \citet{einstein_neue_1906} limit\citep{ando_dynamic_2013}, 
\begin{equation}
	\eta(\phi) = \mu\left(1+\frac{5\phi}{2}\right);
	\label{einstein-3d}
\end{equation}
effective viscosities are discussed more in \cref{sec:viscosity-measurements}.

Replacing $\Minfty$ with this far-field limit gives a considerable time-saving: it reduces an $\mathcal{O}(N^2)$ calculation of a dense matrix followed by its inversion, $\mathcal{O}(N^3)$, with filling a diagonal, $\mathcal{O}(N)$; and it means that the remaining grand resistance matrix,
\begin{equation}
	\R_\text{LH} = (\Minfty_{\text{LH}})^{-1} + \Rtbexact - \Rtbinfty_{\text{LH}},
	\label{grand-resistance-matrix-LH}
\end{equation}
is sparse, but at the expense of accuracy. 

Note that in \cref{grand-resistance-matrix-LH}, the first and third terms do not cancel for systems of more than two spheres, even if the viscosity term in the first matrix is not $\phi$-dependent. To see this, recall that the purpose of $\Rtbinfty$ (and hence $\Rtbinfty_\text{LH}$) is to remove inadvertently-included mobility interactions from $\Rtbexact$. The elements of $\Rtbinfty_{\text{LH}}$ depend on the number of close pairs of particles, whereas $\Minfty_{\text{LH}}$ simply consists of a self-term for each particle. Thus we are left with the first term representing unbounded Stokes flow for each particle, and the following two terms representing pairwise lubrication.

The accuracy of the LH approximation for some $n$-disperse non-Brownian systems will be tested here. The accuracy of the approximation for modelling diffusion in dense, polydisperse, Brownian suspensions was examined in \citet{ando_dynamic_2013}. They found that particle diffusion constants were broadly accurate but that intermolecular dynamical correlations were significantly underestimated. 

We first show that viscosity measurements taken in monodisperse, periodic systems under continuous shear, are qualitatively unaffected by switching from SD to the LH approximation. There is a systematic error in the stresses which results in reduced measurements (up to 20\%) at higher concentrations, but otherwise the system behaviour broadly matches the full SD readings.

In systems where the particles are given external forces, rather than simply moving due to an imposed shear, we begin to find anomalous results for the LH approximation. In particular, we find that we have to be careful about the application of the lubrication forces for bidisperse suspensions. We will show how the default application leads to unphysical results. 

To do this, we examine the motion produced by the LH method of up to five close spheres in simple test cases, both monodisperse and bidisperse. These simple test cases demonstrate the mechanism by which these unphysical results are predicted by the LH approximation for much larger bidisperse suspensions.

In particular, we find that under an external force on a large particle, small particles `bunch up' behind the large particle. In an adaptive timestepping regime, the resultant decrease in particle separations would require reducing the timestep at every timestep. In the test cases, as we will see for viscosity measurements in larger suspensions, we find motion driven by an applied shear to be mostly unaffected, with a small accuracy loss. In the finite test cases, the local concentration is difficult to define, so we use the unaltered solvent viscosity, $\mu$. The lubrication critical radius, $r^*$, is set to be as large as necessary so that all particle pairs are included.

It is first worth noting that for two spheres, SD and LH methods will produce the same result, since the true $\Minfty$ matrix (term 1 in the right-hand side of \cref{grand-resistance-matrix}) will match the sum over all pairwise mobility matrices (term 3 in the same equation). At higher numbers of spheres, a discrepancy grows.

\section{Viscosity measurements}\label{sec:viscosity-measurements}
\begin{figure}
\centering
\includegraphics[width=0.9\textwidth]{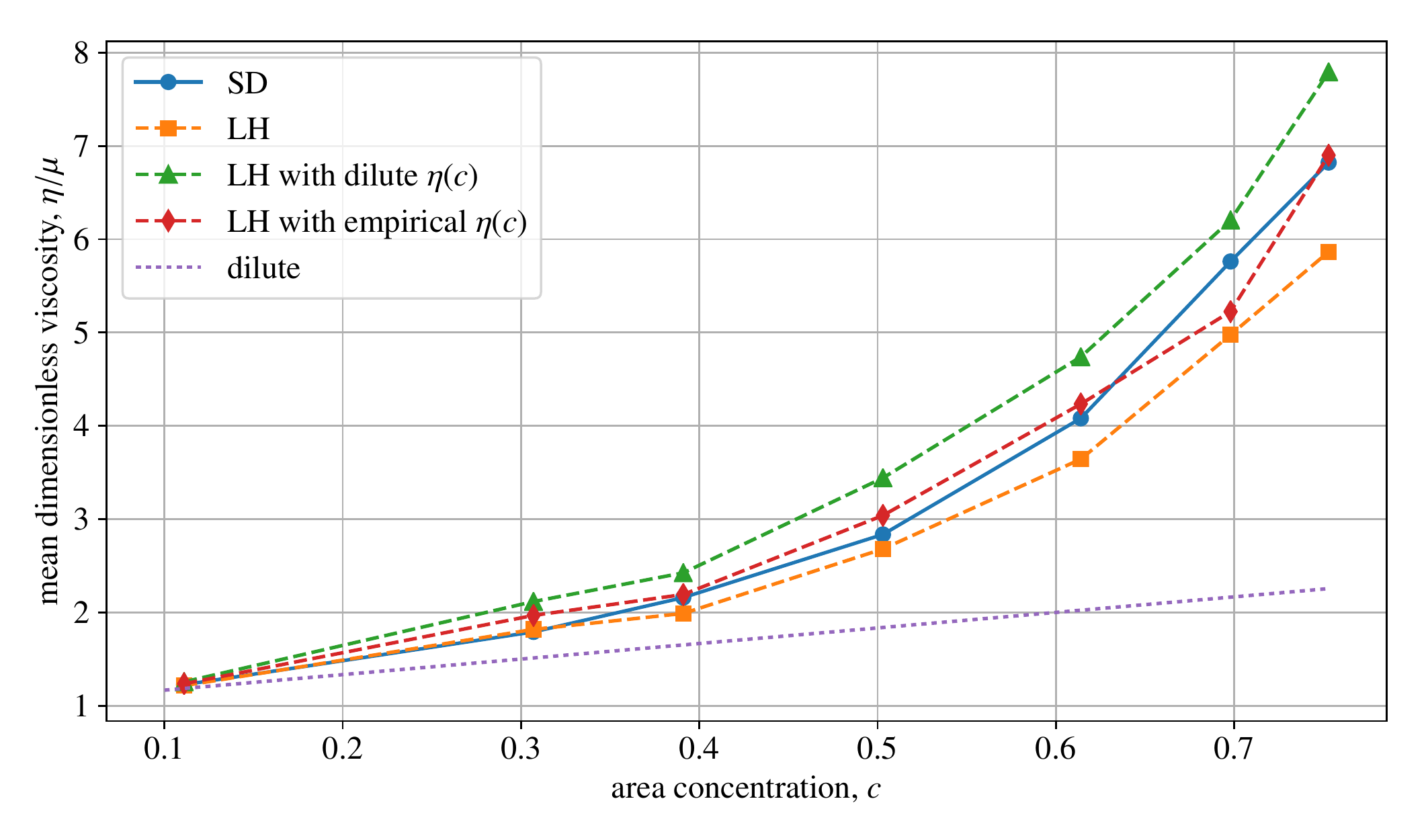}
\caption{The viscosity of a monodisperse monolayer suspension of varying concentration, $c$, in a periodic system is measured with a full SD simulation \mbox{(---)} and with the LH approximation \mbox{($--$)}. The LH approximation is performed with the standard background viscosity, $\eta=\mu$ (squares); with an effective viscosity given by the dilute limit, $\eta=\mu(1+5c/3)$ (triangles); and with an empirical effective viscosity, chosen to match the SD data, $\eta=\mu(1+5c/6)$ (diamonds). The dilute limit is also shown \mbox{($\cdots$)}. The periodic box has side length $15$ particle radii and simulations were performed of initially randomly-positioned particles in constant shear, $\dot\gamma=1$, with 800 RK4 timesteps of size $\mathrm{\Delta} t = 0.005$.}
\label{pic:viscosity-monodisp-minfonoff}
\end{figure}
We first measure the viscosity of a periodic monodisperse single plane of spheres (a monolayer) undergoing continuous shear, using full SD and using the LH approximation. The viscosity contribution from the particles is calculated in the simulations from the particle stresslets, $\t{S}$. The effective viscosity for a three-dimensional simulation with solid volume fraction $\phi$ in a volume $V$, undergoing shear at a constant rate, $\dot\gamma$, in the $xy$-plane is given by
\begin{equation}
	\eta = \mu + \frac{1}{\dot\gamma V}\sum_{\alpha}S^\alpha_{xy},
\end{equation}
where the summation is over all particles $\alpha$. For well-separated spheres, the stresslet is given by
\begin{equation}
	\t{S}^\alpha = \frac{20}{3}\pi\mu a^3 \t{E}^\infty \implies S^\alpha_{xy} = \frac{10}{3}\pi\mu a^3 \dot\gamma,
\end{equation}
which leads to the Einstein relation, \cref{einstein-3d}. In a monolayer, following the convention of \citet{brady_rheology_1985} to take the nominal layer depth as the particle diameter, $2a$, the equivalent dilute-limit effective viscosity for an area fraction, $c$, is
\begin{equation}
	\eta(c) = \mu\left(1+\frac{5c}{3}\right).
	\label{eta-c-anomolous}
\end{equation}

Particles are given a contact force as described in \citet{townsend_frictional_2017}. For pairs of approaching particles, this contact force acts in the direction normal to the particle surfaces to exactly stop the approach once the particle surfaces become sufficiently close (here, $10^{-2}a$). No other forces, such as tangential friction forces or repulsion forces, are imposed on the particles. The system is then placed under continuous shear, $\dot\gamma=1$, and the viscosity is taken from the average of three shear cycles, measured after two shear units have passed. This gives time for the system to equilibrate.

\Cref{pic:viscosity-monodisp-minfonoff} shows the recorded viscosity at different concentrations for SD and LH. We see very good agreement at low concentrations, but find that at higher concentrations, the viscosity readings are underestimated with the LH approximation, up to 20\%. The graph shape is qualitatively right, however. The dilute limit, \cref{eta-c-anomolous}, is shown on the graph for comparison. 

The suggestion in \citet{ando_dynamic_2013} (for fully 3D suspensions) of changing the viscosity term, $\mu$, in \cref{minfty-far-field-limit}, to the dilute effective viscosity, is also shown on the graph. We find that it overestimates the viscosity by about 20\%, suggesting that perhaps this effective viscosity switch does not work particularly well in monolayers. This may be due to the different sphere-packing properties in 2D and 3D. Instead, we find that an empirical effective viscosity of $\eta(c)=\mu(1+5c/6)$ gives better agreement. 
Recalling that the $5c/3$ term in \cref{eta-c-anomolous} is derived from the assumption that the monolayer has `effective depth' $2a$, the empirical effective viscosity therefore suggests an `effective depth' of $4a$ in order to scale in the same way as the fully 3D solution, i.e., that the dilute limit is an appropriate effective viscosity.

A very similar comparison experiment for a fully 3D suspension is found in \citet[fig.~2.16]{bybee_hydrodynamic_2009}---where their `fast lubrication dynamics' is the same as our LH but with a further approximation to $\Rtbexact$---and draws the same conclusion.%
\section{Monodisperse test cases}%
\begin{figure}
\centering
\includegraphics[scale=1.25]{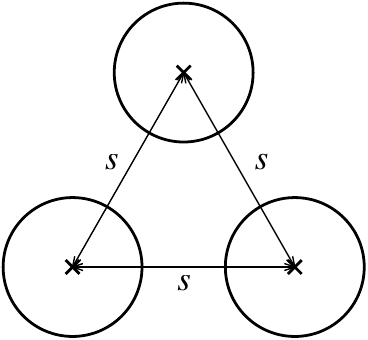}
\caption{Three identical spheres are arranged in an equilateral triangle with side length $s$, and are given a force perpendicular to the plane in which they lie.}
\label{pic:3inatriangle}
\end{figure}%
\begin{figure}
\centering
\includegraphics[scale=1.05]{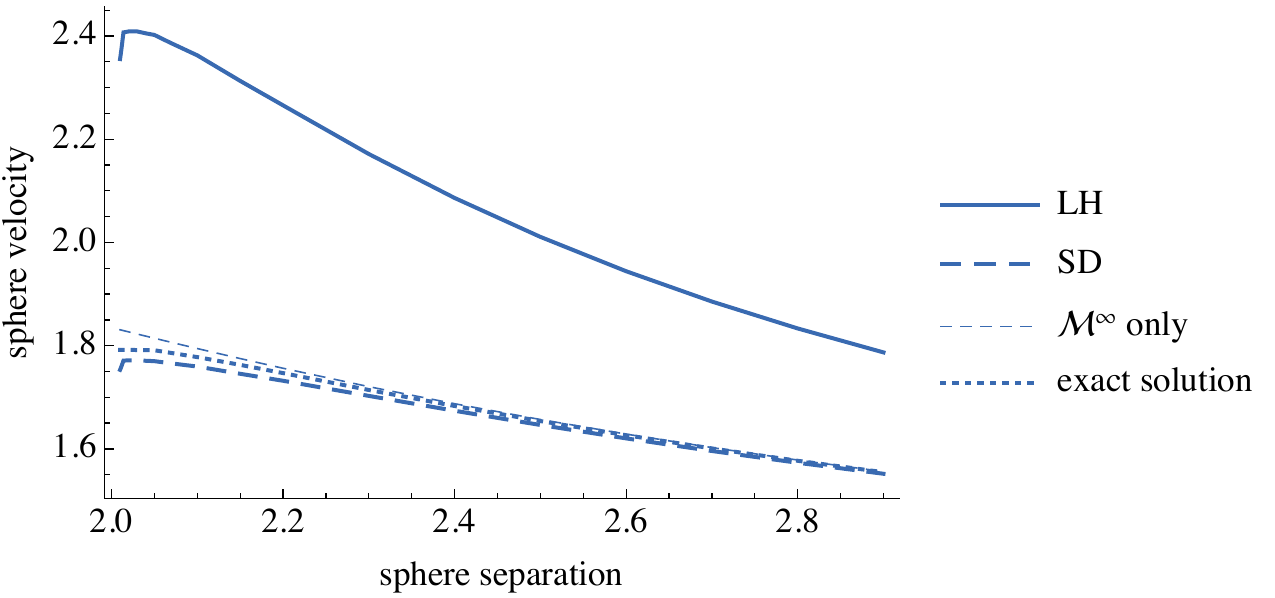}
\caption{Velocity of an equilateral triangle of spheres given identical forces perpendicular to their plane. Full SD matches the exact solution from the \citet{wilson_stokes_2013} 3-sphere code well apart from at values very close to $s=2$, but LH overestimates the velocity by up to 30\%.}
\label{pic:3sphere-triangle-comparison}
\end{figure}
We illustrate the discrepancy between using the full SD grand resistance matrix, \cref{grand-resistance-matrix}, and the simplified far-field LH form, \cref{grand-resistance-matrix-LH}, with a setup from \citet{wilson_stokes_2013}: three identical spheres of radius $a$, arranged in an equilateral triangle with a given side length (see \cref{pic:3inatriangle}). All three spheres are then given a force of $6\pi\mu a$ perpendicular to the plane of the spheres. \Cref{pic:3sphere-triangle-comparison} shows the resultant sphere velocities for both cases, and compares it with the true three-sphere velocity.

In agreement with fig.~2 in \citet{wilson_stokes_2013}, SD matches the exact 3-sphere solution for all separations well, with the largest error (2\%) at very close sphere separations. However, LH shows much worse results, overestimating the velocity by up to 30\% at the smallest separations. The results are considerably worse than those from a run with the long-range mobility matrix $\Minfty$ enabled but the lubrication matrices $\Rtbexact$ and $\Rtbinfty$ disabled (`$\Minfty$ only'): this has an error of at most 5\%. Finally, at high separations, all solutions converge.

\begin{figure}
\centering
\includegraphics[scale=1.25]{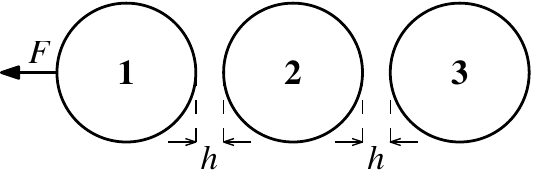}
\caption{Three spheres are aligned in a row, with their surfaces separated by an equal distance $h$. The first sphere is then given a force directly away from the other spheres.}
\label{pic:3inarow-monodisperse}
\end{figure}
\begin{figure}
\centering
\includegraphics[scale=0.7]{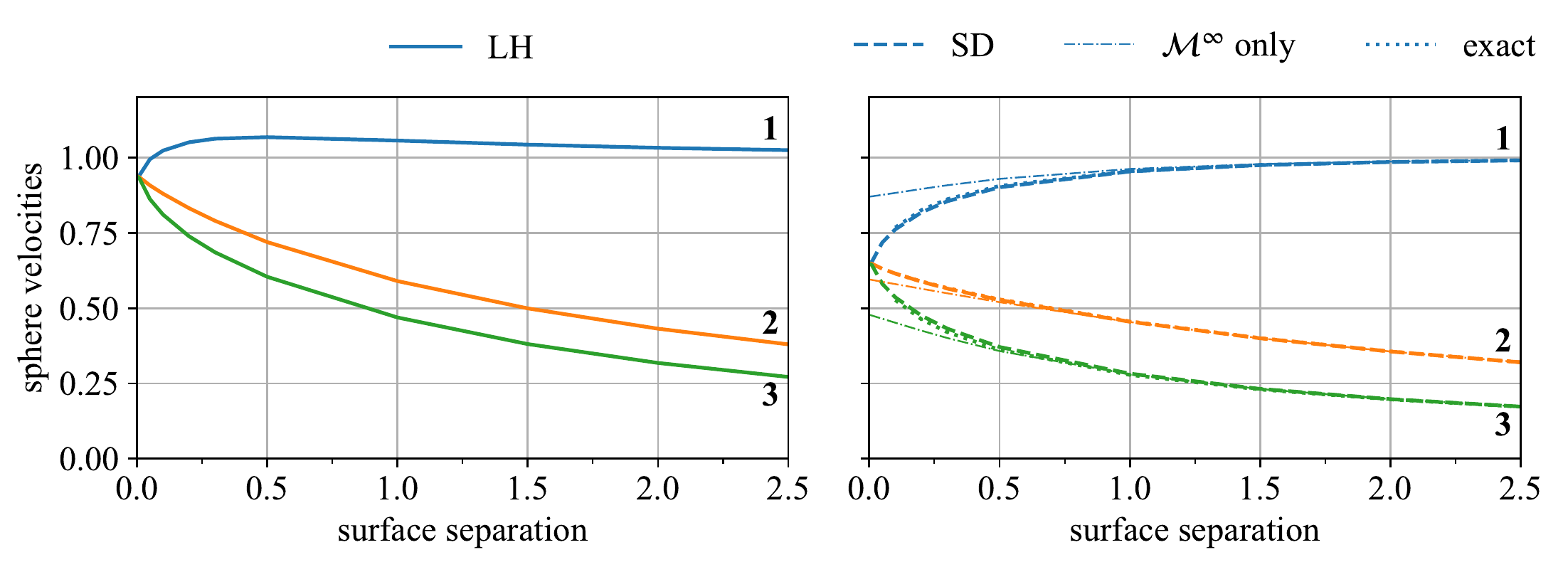}
\caption{Velocities of three identical particles, aligned in a row with a given, equal surface separation, as in \cref{pic:3inarow-monodisperse}. Particle \textbf{1} is given a force directly away from particles \textbf{2} and \textbf{3}, and the velocities are measured with LH (left, ---) and SD (right, $--$). The exact solution from \citet{wilson_stokes_2013} (right, $\cdots$) is also shown, as well as with $\Minfty$ only (right, $-\!\cdot\!-$).}
\label{pic:3inrow-monodisperse-rc20k}
\end{figure}
We are now going to consider a setup of three identical, linearly aligned particles of radius $a$, as illustrated in \cref{pic:3inarow-monodisperse}. The first particle is given a force of $6\pi\mu a$ directly away from the other particles, and the velocities produced with both SD and LH are recorded in \cref{pic:3inrow-monodisperse-rc20k}. We see a similar phenomenon as before: LH results in velocities for all three particles which have a similar profile shape, but whereas they converge to the exact result at high separations, at the smallest separations the readings are up to 45\% larger. Once again this is worse than ignoring lubrication completely (`$\Minfty$ only'), which has a maximum error of 34\%. 
\section{Bidisperse linear test cases}\label{bidisperse-linear}
Although inaccurate, the 30\%--45\% increase in velocity seen in the monodisperse test cases is still qualitatively feasible. Since the shapes of the velocity profiles are similar, in a concentrated suspension, having many more lubrication forces, it can be argued that such local effects might `average out' and would be mitigated in a concentrated suspension by use of the modified effective viscosity, $\mu(\phi)$. With bidisperse suspensions, however, we begin to see unphysical behaviour with the LH approximation. 

\begin{figure}
\centering
\includegraphics[scale=1.25]{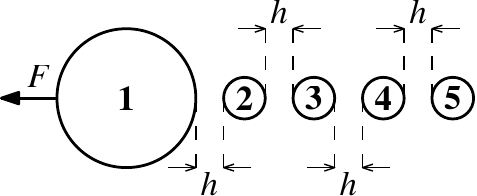}
\caption{A large sphere and a tail of smaller spheres are aligned in a row, with their surfaces separated by an equal distance $h$. The large sphere is then given a force directly away from the smaller spheres in our test cases.}
\label{pic:5inarow}
\end{figure}

\begin{figure}
\centering
\includegraphics[scale=0.7]{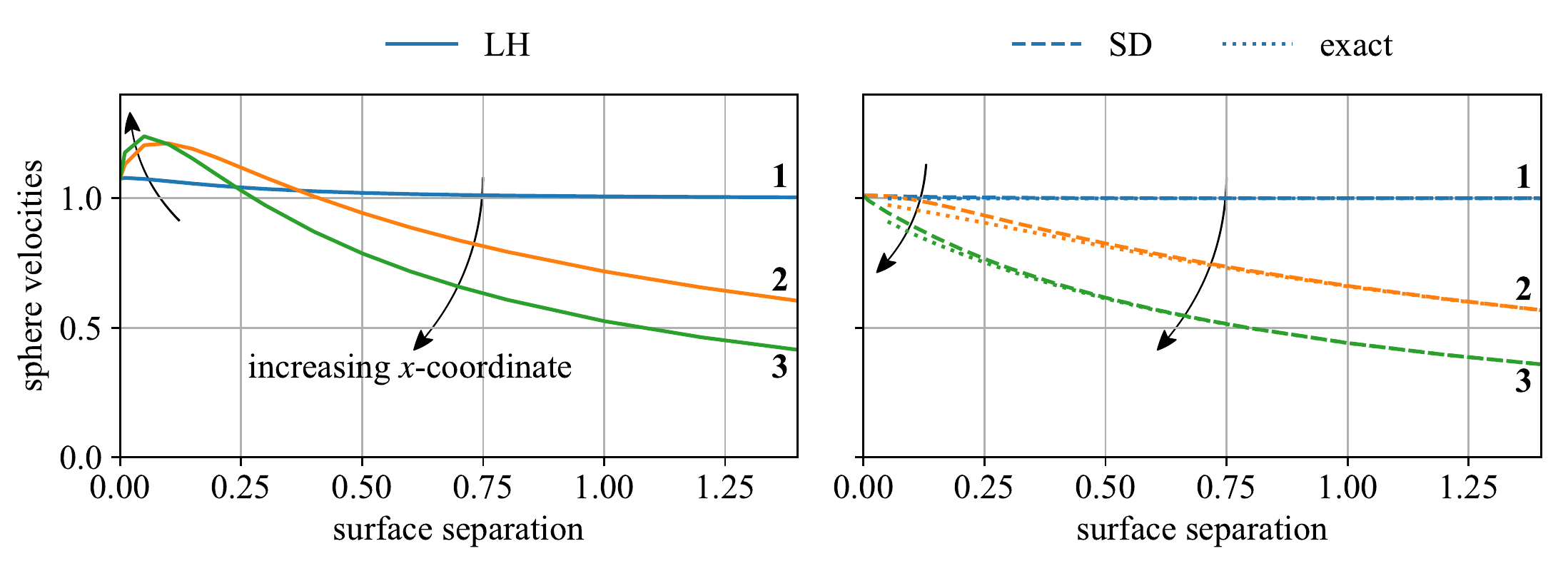}
\caption{Velocities of one large (radius $a$) and two small particles (radius $a/10$), aligned in a row with a given, equal surface separation, i.e., as in \cref{pic:5inarow} but with a shorter tail. The large particle, \textbf{1}, is given a force directly away from small particles \textbf{2} and \textbf{3}, and the velocities are measured with LH (left, ---) and SD (right, $--$). The exact solution from \citet{wilson_stokes_2013} ($\cdots$\!) is also shown. The arrows point towards the tail of the row of spheres, i.e.\ in the increasing $x$-direction.}
\label{pic:3inrow-bidisperse-rc20k}
\end{figure}

This time consider a setup of linearly aligned particles, similar to the last one, but with one large particle (of radius $a$) and two small (of radius $a/10$) particles, as illustrated in \cref{pic:5inarow} but with a shorter tail. The large particle is given a force of $6\pi\mu a$ directly away from the smaller particles and the velocities produced by SD and LH are shown in \cref{pic:3inrow-bidisperse-rc20k}. 

For this setup, the velocity profiles for SD and LH no longer have the same shape. Still, at large surface separations we find convergence of the LH velocities to the exact result (provided by \citet{wilson_stokes_2013}). Full SD agrees well throughout with the exact result, with errors of no more than 4\% for the furthest sphere at small surface separations. However, at these close surface separations, we find the unphysical result of the small particles travelling \emph{faster} than the sphere with the force on it. Furthermore, the small particles travel even faster the further away from the large particle they are, leading to `bunching'. This effect gives rise to particles approaching each other unphysically at the end of the tail, as they `chase' the lead particle too quickly, causing potential numerical instabilities at small timesteps. 

\begin{figure}
\centering
(a)
\includegraphics[scale=0.7,valign=t]{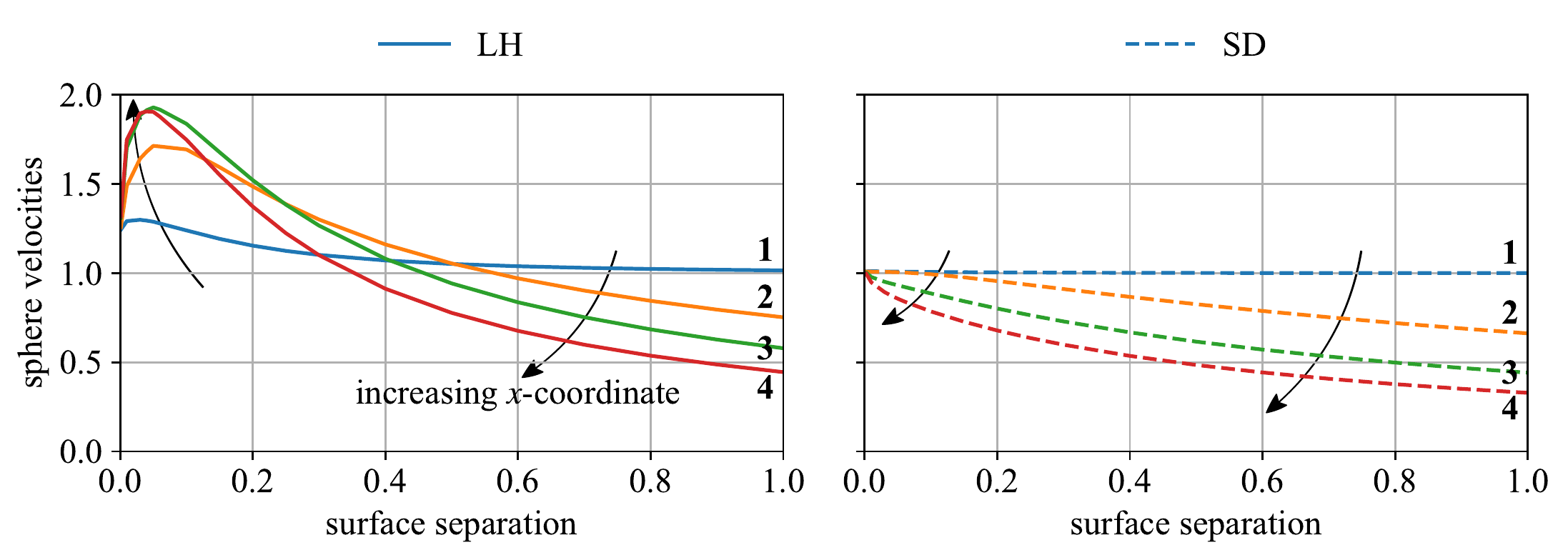}
(b)
\includegraphics[scale=0.7,valign=t]{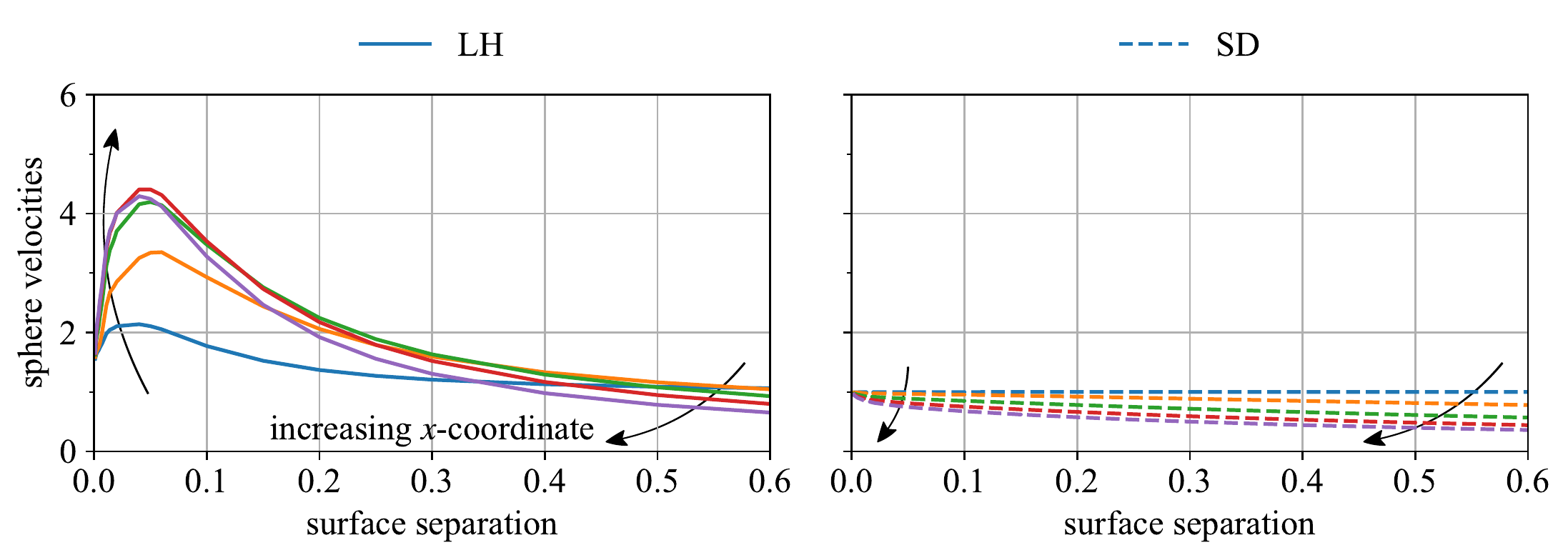}
\caption{\textbf{(a)} Velocities of one large (radius $a$) and three small particles (radius $a/10$), aligned in a row with a given, equal surface separation. The large particle, \textbf{1}, is given a force directly away from the small particles \textbf{2}--\textbf{4}, and the velocities are measured with LH (left, ---) and SD (right, $--$). The arrows point towards the tail of the row of spheres, i.e.\ in the increasing $x$-direction. \textbf{(b)} Same but with one large and four small particles.}
\label{pic:4inrow-bidisperse-rc20k}
\end{figure}
This result is amplified as the tail length increases. \Cref{pic:4inrow-bidisperse-rc20k} shows velocities for tails with three and four small particles. In the latter case, we find velocities of the small particles which are measured to be over five times larger with LH than with SD. 

\begin{figure}
\centering
\includegraphics[scale=0.7]{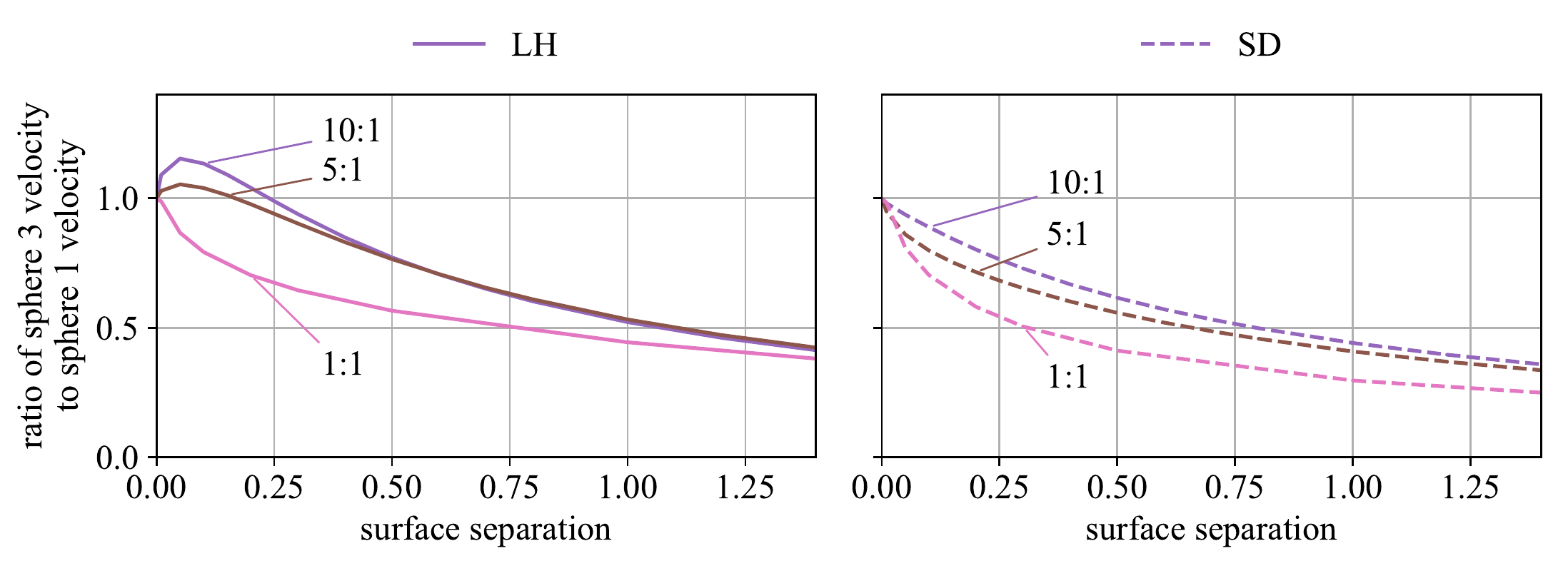}
\caption{Ratio of the velocity of particle 3 to particle 1 in tests of one large particle and two smaller particles, aligned in a row with a given, equal surface separation, i.e., as in \cref{pic:5inarow} but with a shorter tail. The size ratio $a_1:a_3$ is given in each case. The large particle, 1, is given a force directly away from small particles 2 and 3, and the velocities are measured with LH (left, ---) and SD (right, $--$). The unphysical effects under LH, indicated by any values of the velocity ratio large than $1$, increase with size ratio.}
\label{pic:3inrow-size-ratios}
\end{figure}
The unphysical effects also grow as the size ratio $a_\text{large}/a_\text{small}$ increases. \Cref{pic:3inrow-size-ratios} demonstrates this with three different size ratios. In each case, a three-particle system---one large, two small, as in \cref{pic:5inarow} but with a shorter tail---is considered. A force of $6\pi\mu a_\text{large}$ is given to the first particle, directly away from the tail, and the ratio of sphere 3's velocity to sphere 1's velocity is measured for different initial surface separations. Recalling that any value of this velocity ratio larger than $1$ indicates the unphysical behaviour, we see that increasing the size ratio leads to growth of the anomalous effect.

\begin{figure}
\centering
\includegraphics[scale=1.25]{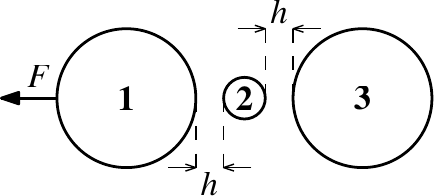}
\caption{A large sphere, a small sphere, and another large sphere are aligned in a row, with their surfaces separated by an equal distance $h$. The first large sphere is then given a force directly away from the other spheres in our test cases.}
\label{pic:big-small-big}
\end{figure}

\begin{figure}
\centering
\includegraphics[scale=0.7]{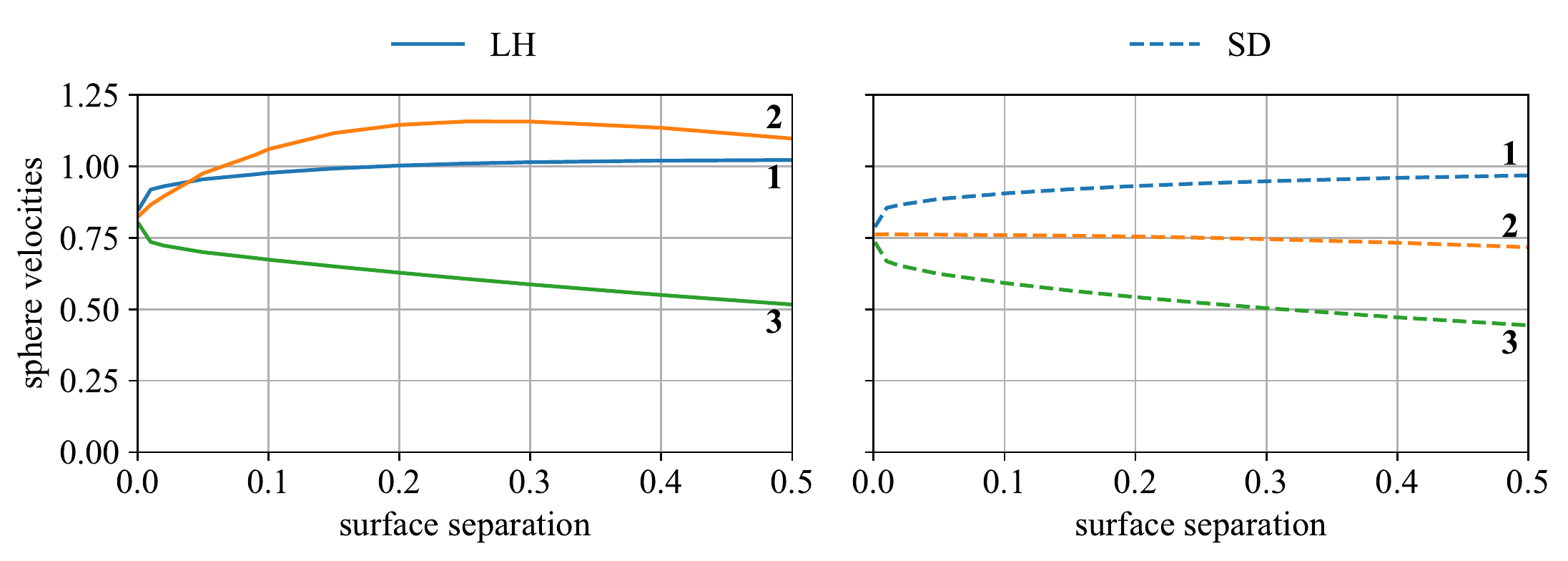}
\caption{Velocities of one large (radius $a$), one small (radius $a/10$), and another large particle (radius $a$), aligned in a row with a given, equal surface separation, i.e., as in \cref{pic:big-small-big}. The large particle, \textbf{1}, is given a force directly away from small particle \textbf{2} and large particle \textbf{3}, and the velocities are measured with LH (left, ---) and SD (right, $--$).}
\label{pic:3inrow-big-small-big-rc20k}
\end{figure}

Finally, we look at a large--small--large test case, as in \cref{pic:big-small-big}. Once again, the first large particle is given a force of $6\pi\mu a$ directly away from the tail, and the velocities of the particles for different initial separations, under LH and SD, are recorded in \cref{pic:3inrow-big-small-big-rc20k}. This time we see that the large particles under LH behave broadly appropriately (as we saw in the monodisperse case), but the small particle once again travels faster than the first.

We only find this bunching effect with applied external forces. Placing the same system in an external shear produces an acceptable error between the SD and LH simulations, similar to the monodisperse case. 
%

\section{Mechanism}
The mechanism we propose for the bunching behaviour described in this section comes from the reach of the lubrication forces. These forces have a stronger effect on the small spheres than on the large one, and the setting of a critical radius, \cref{r-star}, means that we can find multiple sets of lubrication interactions on the small spheres. In particular, with a large particle at the head and multiple small particles in the tail, the last small sphere in the tail feels all of these forces pulling it in the same direction, giving it a larger velocity than the others. 

In reality, the small spheres in between provide screening against this effect. However, this is exactly what the full long-range mobility matrix $\Minfty$ captures and is what we have lost: inverting this matrix is equivalent to summing reflected interactions among all particles.\citep{durlofsky_dynamic_1987,brady_stokesian_1988} Errors caused by the omission of screening have also been seen in spectral convergence studies of interacting spheres in bounded domains,\citep{navardi_general_2013,navardi_stokesian_2015} where the replacement of the long-range mobility matrix $\Minfty$ with its far-field limit in the LH approximation is equivalent to reducing the multipolar order of the calculation from 2 to 1. That the effect is worse when the size ratio is increased, as seen in \cref{pic:3inrow-size-ratios}, further demonstrates the importance of the lost screening.

\section{Proposed solution}
The question for LH is which lubrication interactions to enforce, given that the setting of a critical radius leads to unphysical results. The large--small--large case of \cref{pic:big-small-big} provides a good place to examine possible solutions. Given that this case is symmetric and all particles need to feel some lubrication interaction in order to move, there are only two choices. Either all pairs of particles interact, or only the nearest neighbours interact. 

\begin{figure}
\centering
\includegraphics[scale=1]{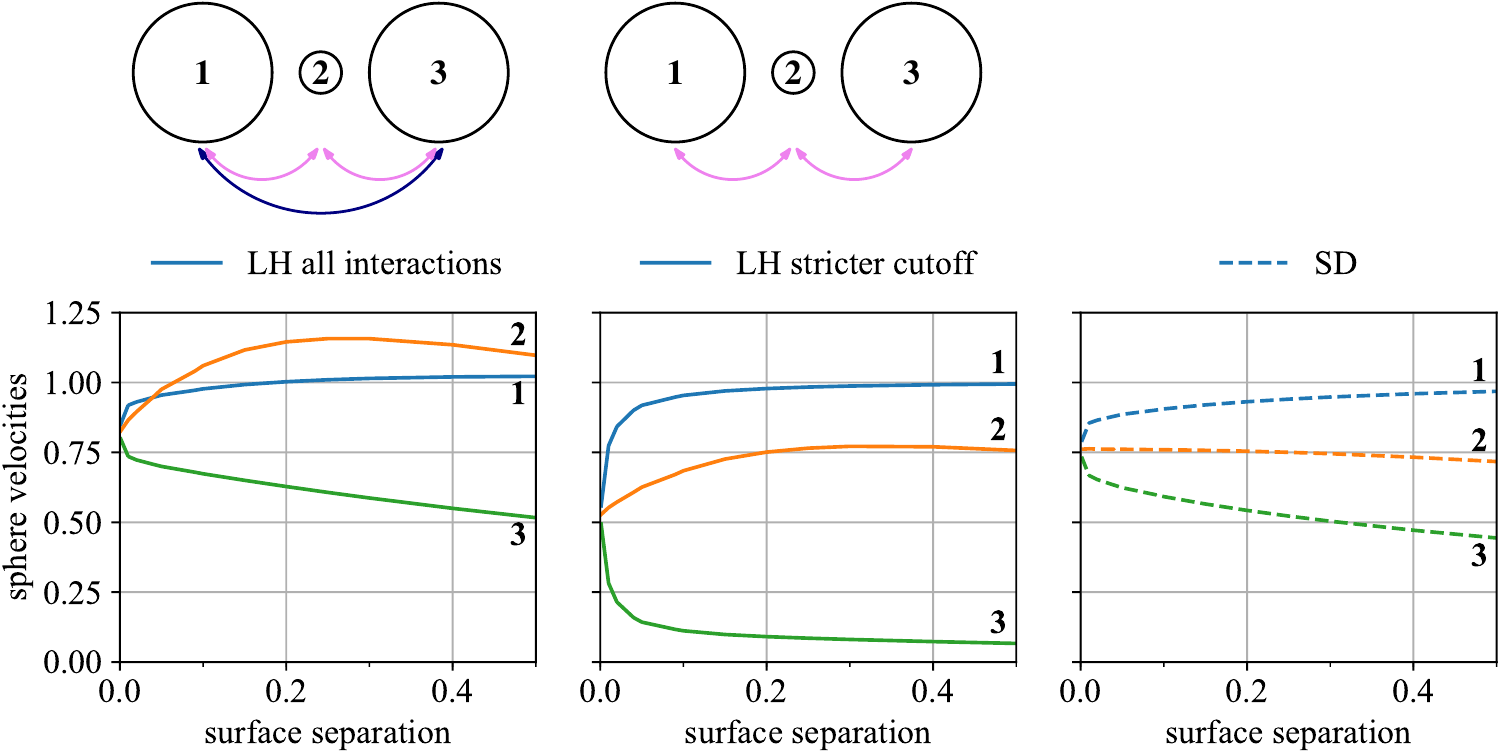}
\caption{Sphere velocities from the test case in \cref{pic:big-small-big} under LH with interactions between different particles enabled, compared to velocities with SD. Far left: LH with all interactions. Centre: LH with interactions only on pairs where other particles cannot pass between them, which is equivalent to the stricter global cutoff \cref{r-star-fixed}.}
\label{pic:big-small-big-options}
\end{figure}
\Cref{pic:big-small-big-options} explores these options. From this, we can see that the only viable option is to implement lubrication only on pairs where another particle cannot pass between them (the centre graph). This is equivalent to setting a global cutoff so that we only implement lubrication between pairs of particles of sizes $a_1, a_2$ when their centre-to-centre separation distance, $s$, is
\begin{equation}
s < a_1 + a_2 + 2a_{\text{small}},
\label{r-star-fixed}
\end{equation}
where $a_\text{small}$ is the radius of the smallest particle in the system. For systems of identical spheres, this is equivalent to the conventional cutoff, \cref{r-star}.

We find that with the other test cases as well, this approach to lubrication---effectively, aggressive screening---is the only option which consistently avoids the unphysical behaviour the LH method can produce. As can be seen in the figure, however, the method clearly underpredicts the speed of the particles, particularly the farthest particle. In a system with a large number of forced particles, it is plausible that this reduced effect on neighbours is small compared to the driving force on each particle, but even this can be quite significantly reduced: the figure shows a 30\% reduction of the lead particle speed at the closest separation compared to SD.

\section{Conclusion}
The efficacy of replacing the long-range mobility matrix $\Minfty$ in Stokesian Dynamics with its far-field form has been tested. We find it to be appropriate in shear-driven, periodic suspensions, but for bidisperse suspensions where the particles are exposed to an external force, it can produce errors if we are not careful about how we apply the lubrication forces. 

For monodisperse suspensions under applied force or bidisperse suspensions under applied shear, these errors are large for small separations but affect all the particles equally. However, for bidisperse particles under applied force, the error disproportionately affects the smaller particles, giving them unphysical velocities which can lead to particles approaching each other too quickly. This cannot be mitigated by the choice of numerical method. That the effect is greater with increasing numbers of particles is particularly concerning. We suggest, therefore, that methods involving this lubrication hydrodynamics simplification should therefore be used with caution when applying external forces to bidisperse suspensions. This effect can be mitigated by enabling lubrication only between pairs closer than the stricter cutoff, \cref{r-star-fixed}.

\bibliography{chap-stokesian-dynamics-references}{}

\end{document}